\documentclass[aps,twocolumn,pra,superscriptaddress,showpacs,tightenlines,floatfix]{revtex4-1}

\usepackage{graphicx,amssymb}
\usepackage{amsmath}
\def \etal{{\em et al.}}

\begin{document}

\title{Measuring nonclassical correlations of two-photon states}

\author{Karol Bartkiewicz}
\email{bartkiewicz@jointlab.upol.cz}
\author{Karel Lemr}
\author{Anton\'{i}n \v{C}ernoch}
\author{Jan Soubusta}

\affiliation{RCPTM, Joint
Laboratory of Optics of Palack\'y University and Institute of
Physics of Academy of Sciences of the Czech Republic, 17.
listopadu 12, 772 07 Olomouc, Czech Republic }

\date{\today}

\begin{abstract}

The threshold between classical and nonclassical two-qubit states is drawn at the place when these states can no longer be described by classical correlations, i.e., quantum discord or entanglement appear. However, to check if the correlations are classical (in terms of quantum discord and entanglement) it is sufficient to witness the lack of quantum discord because its zero value implies the lack of entanglement.  We explain how the indicator of quantum discord introduced by Girolami and Adesso [Phys. Rev. Lett. \textbf{108}, 150403 (2012)] can be practically measured in linear-optical systems using standard beam splitters and photon detectors. We study the efficiency of the setup assuming both ideal and real components and show that the efficiency of the proposed implementation is better than the full two-photon quantum tomography.  Thus, we demonstrate that a class of experiments previously available on NMR platform can be implemented in optical systems.  

\end{abstract}

\maketitle

\section{Introduction}

The threshold between the classical and quantum world has fascinated physicists since the discovery of quantum phenomena and realizing how different they are from our everyday experience. One of the prominent examples of quantum behavior is the nonlocality leading to violation of Bell's inequalities \cite{Bell1964,Clauser1969}. For two-level systems there is no nonlocality without quantum entanglement, but the opposite can be true \cite{Werner1989}. Quantum entanglement plays an important role in quantum information processing \cite{Horodecki2009}. However, the entanglement is not the only type of non-classical correlations.  As described by  Ollivier and Zurek \cite{Olivier2001} the nonclassical correlations can be associated  with \textit{quantum discord}. Quantum discord (QD) is useful in many ways including quantum information processing or detection of quantum phase transitions, especially in the cases when the entanglement fails to grasp this phenomenon \cite{Modi2012}. Moreover, it was demonstrated that only classical correlations can be broadcast locally \cite{Piani2008}. All of these features of quantum discord motivate the quest for developing tools for detecting and quantifying it. Nevertheless,  there were only a few experimental implementations of witnesses of nonclassical correlations, or \textit{nonclassicality witnesses} (NWs), in discrete-variable systems. Two of them were implemented in nuclear magnetic resonance systems \cite{Passante2011,Auccaise2011} and one using linear optics \cite{Aguilar2012}, however these witnesses were not universal.  At this point, we should stress that detecting purely classical correlations is a difficult problem since it involves solving optimization problem over a nonconvex set of classical states. Thus, the problem of detecting classical correlations is harder that detection of entanglement. Moreover, any NW should be nonlinear \cite{Rahimi2010}. For those reasons the NWs \cite{Rahimi2010,Maziero2012,Aguilar2012} are usually non-universal. However, Zhang \etal  \cite{Yu2011} demonstrated that finding a universal NW is possible, but the established witness is not suitable for optical implementation. A better suited QD indicator (QDI)  which overcomes the limitations of all the previously developed witnesses is a measure introduced by Girolami and Adesso \cite{Girolami12}. We call it an indicator instead of a witness since in contrast to a typical witness it is universal and on average its value provides a bound on QD.

Let us start with  introducing some basic definitions used throughout our paper. A general two-qubit density matrix $\rho$ can be expressed in the Bloch representation as
\begin{equation}
\rho  =  \frac{1}{4}(I\otimes I+\vec{x}\cdot\vec{\sigma}\otimes I+I\otimes\vec{y}\cdot\vec{\sigma}+\!\!\!\sum \limits_{n,m=1}^{3}T_{nm}\,\sigma_{n}\otimes \sigma _{m}), \label{eq:rho}
\end{equation}
where $\vec{\sigma}=[\sigma_{1},\sigma_{2},\sigma_{3}]$ and matrix $T_{ij}=\mathrm{Tr}[\rho(\sigma_{i}\otimes\sigma_{j})]$ are given in terms of the Pauli matrices, and $x_{i}=\mathrm{Tr[}\rho(\sigma_{i}\otimes I)]$ ($y_{i}=\mathrm{Tr[}\rho(I\otimes\sigma_{i})]$) describe Bloch vector $\vec{x}$ ($\vec{y}$) of the first (second) subsystem, later referred to as $A$ and $B$. Moreover, it is always possible to transform $\rho$ with local unitary operations \cite{Luo2009} so that $T$ becomes a diagonal matrix. 

The state $\rho$ is not entangled (is separable) when it has a positive partial transpose, i.e., is a PPT state (see Peres-Horodecki criterion \cite{Peres1996,Horodecki1996}). The lack of entanglement for a two-qubit system implies, e.g., locality, in terms of violation of the Bell-CHSH inequality \cite{Clauser1969} (for quantitative study see \cite{Horst2013}), and thus it corresponds to classical situation where the measurement outcomes can be explained by a hidden-variable model. However, quantum entanglement is not responsible for all the nonclassical effects. One of the recently celebrated manifestation of quantumness is \textit{quantum discord} \cite{Olivier2001}. The QD is responsible for the difference in conditional quantum information calculated in two ways, where one of them uses the Bayesian rule for calculating probabilities. Therefore, QD quantifies how much conditional quantum probabilities differ from those calculated within classical theory. The QD vanishes if the state fulfills the strong PPT condition \cite{Bylicka2010}, i.e., $\rho$ has to be PPT and its PPT must admit Cholesky decomposition (there are also other so-called nullity conditions -- for review see \cite{Modi2012}).  Thus, if there is no discord, there is no entanglement. However, the reverse does not have to be true. 

There are several ways of quantifying QD. The one for which an analytic formula is known \cite{Dakic2010}  is the so-called \textit{geometric quantum discord} (GQD) quantifying Hilbert-Schmidt distance to the closest non-discordant state. The expression for the GQD reads
\begin{equation}
\label{eq:g_discord}
D_i(\rho)=-\frac{1}{4}\left( \lambda_{\max,i} -\sum_{n=0}^2\lambda_{n,i}\right),
\end{equation}
where $\lambda_{n,i}$ (for  $i=A,B$) stand for eigenvalues of matrix $K_A=\vec{x}\vec{x}^T+TT^T$ or $K_B=\vec{y}\vec{y}^T+TT^T$, where ${}^T$ denotes transposition. The largest  $\lambda_{n,i}$ is denoted as $\lambda_{\max,i}$. Note that $D_i$ is asymmetric. Thus, if $D_A=0$ the state is called classical quantum or if $D_B=0$ the state is quantum-classical. Naturally, there have been attempts of finding an analytic formula for the symmetric GQD, which answers the question about the closest classical-classical state, however this is still an open problem \cite{Modi2012,Miranowicz2012}. 

If $D_A=D_B=0$ the state is classical-classical since it does not exhibit quantum correlations responsible for discord between conditional quantum information calculated in the two above-mentioned ways. In the following sections we show how to experimentally identify states of zero $D_i$ and describe how to perform the experiment within the framework of linear-optics by measuring $Q_i$ introduced in in Ref.~\cite{Girolami12}. The QDI  provides a tight and faithful lower bound ($Q_i=0\Leftrightarrow D_i=0$) for GQD and reads
\begin{equation}
Q_i = \frac{1}{12} \left[ 2M_{1,i} -\sqrt{ 6M_{2,i}-2M_{1,i}^2 }\right]\leq D_i, \label{eq:witnessA}
\end{equation} 
where $M_{n,i}=\sum_{m=0}^2\lambda_{m,i}^n$ for $n=1,2$ are moments of the matrix $K_i$ ($i=A,B$) from Eq.~(\ref{eq:g_discord}), where $\lambda_m$ denotes $m$th eigenvalue of $K_i$. Note that $Q_i$ and $K_i$ are asymmetric, thus $Q_i$ cannot exclusively detect classical-classical states. One of the possible symmetric QDIs is $Q_s=Q_A + Q_B$. Moreover, since  the symmetric geometric discord vanishes $D_s=0$ only if $D_A=D_B=0$ \cite{Miranowicz2012}, for checking if the state exhibits classical-classical correlations $Q_s=Q_A+Q_B$ should be used.

\begin{figure}
\includegraphics[width=8cm]{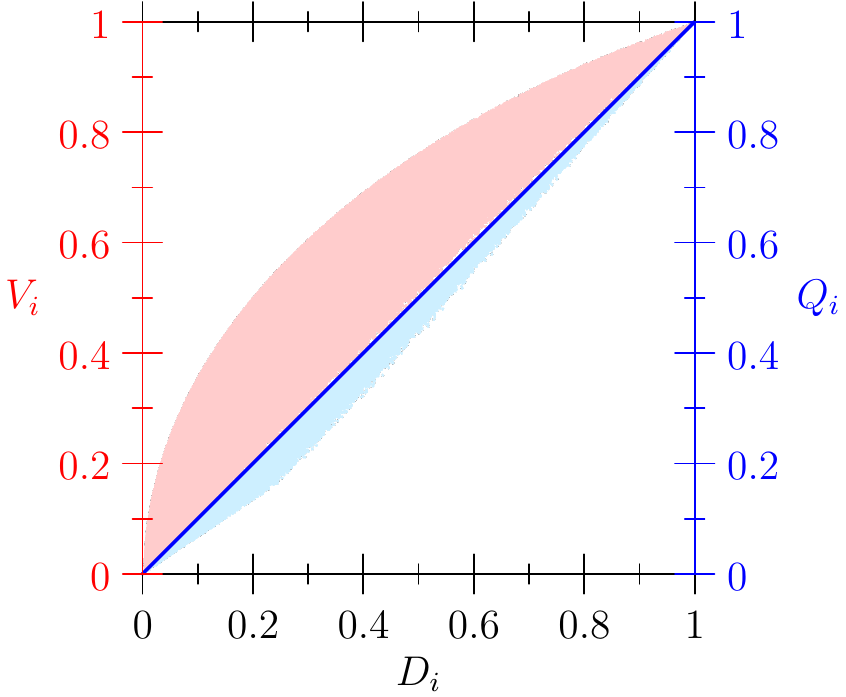}
\caption{\label{fig:1}(Color online)  Monte Carlo simulation outcomes for $10^6$ random two-qubit density matrices. Geometric quantum discord $D_i$ versus normalized quantum discord indicators $Q_i$ and $V_i$. 
}
\end{figure}

Finally, let us note that there are other QDIs than $Q_i$ that are functions of the moments $M_{n,i}$ for $n=1,2$ providing a faithful bound on geometric quantum discord, e.g.,
\begin{equation}
V_i = \sqrt{M_{2,i} - M_{1,i}^2}\geq D_i,
\end{equation}
however, as demonstrated in Fig.~\ref{fig:1}, $Q_i$ provides a better estimation of $D_i$.

\begin{figure}
\includegraphics[scale=0.33]{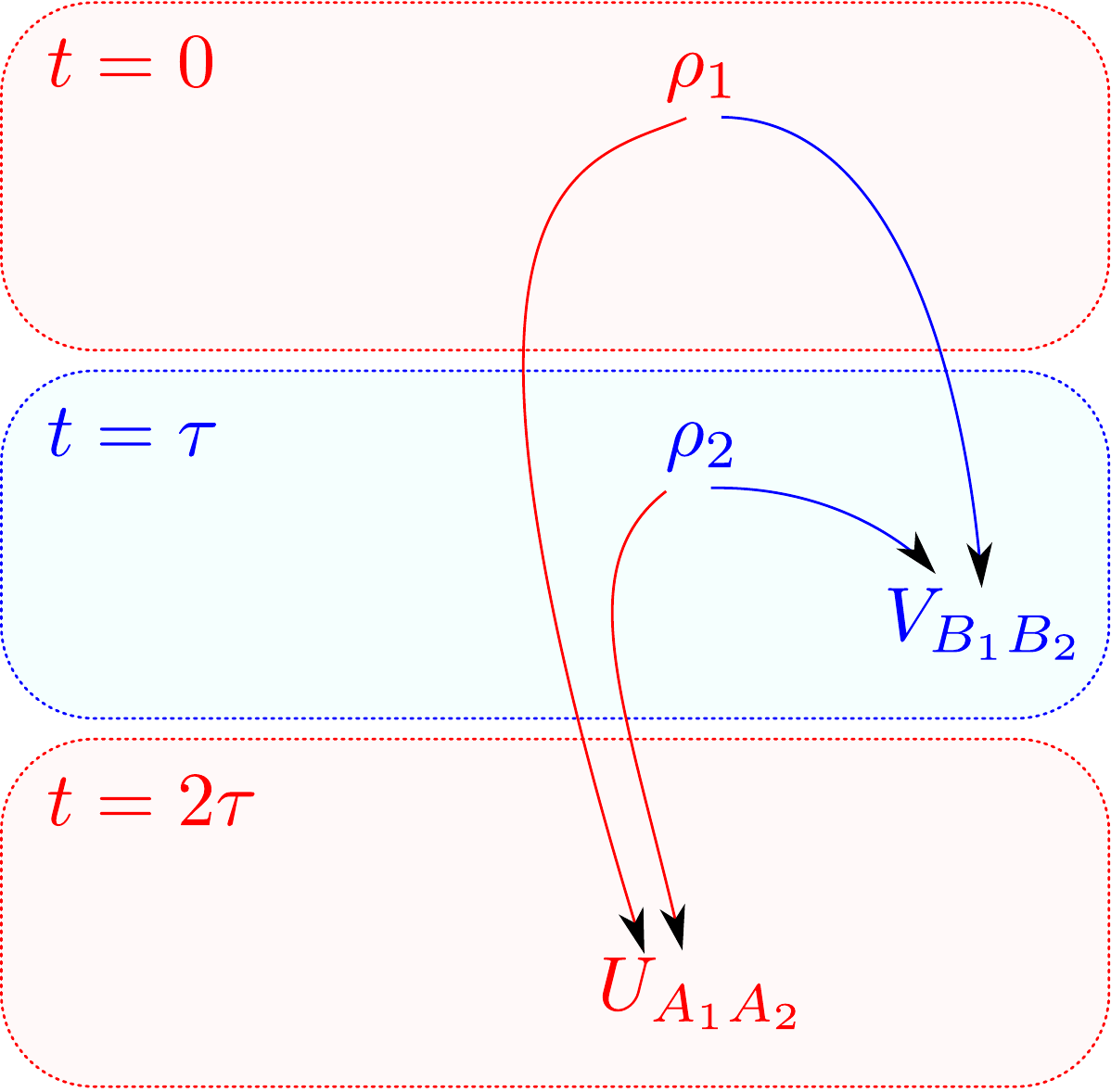}
\includegraphics[scale=0.33]{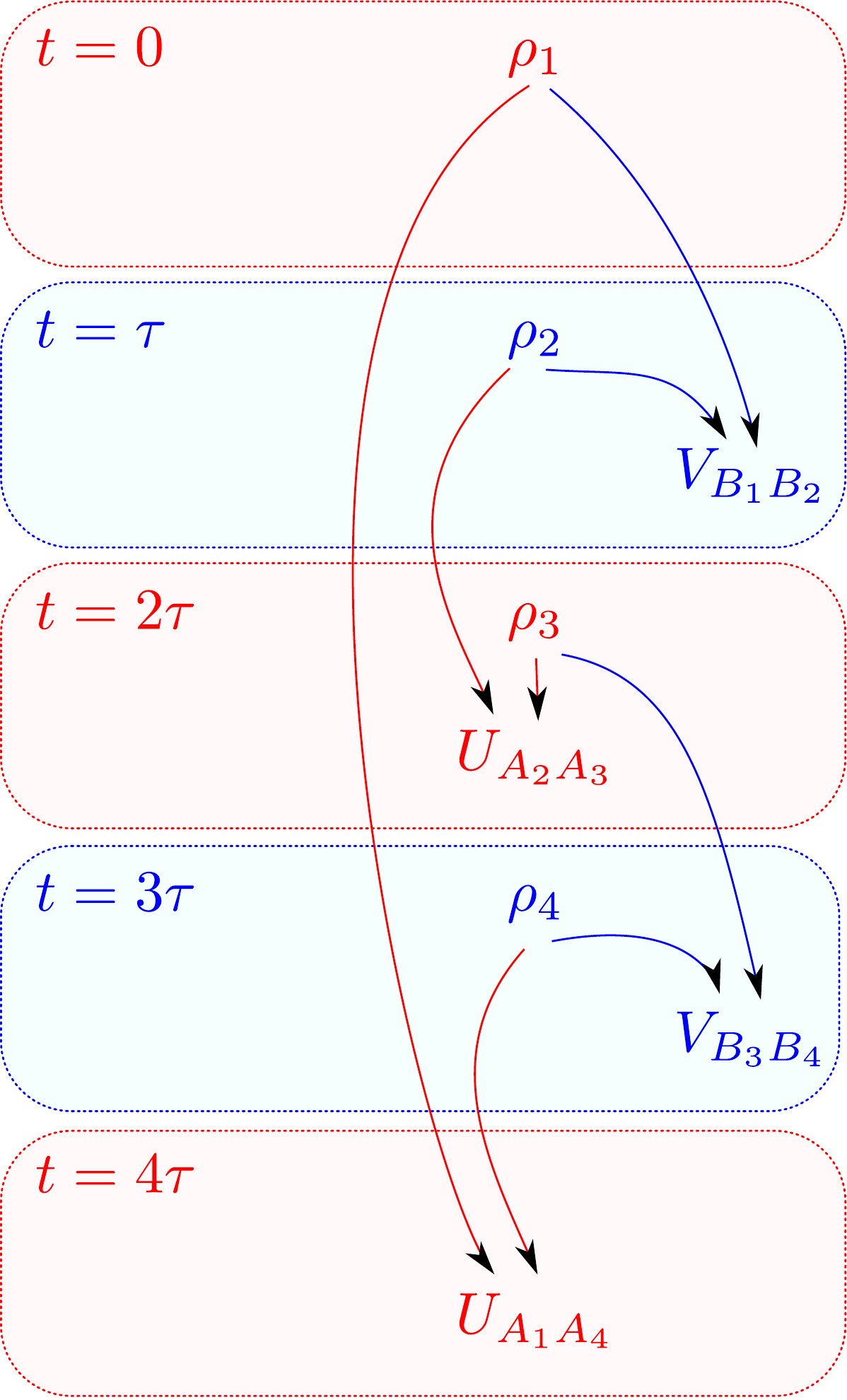}
\caption{\label{fig:2}(Color online) Diagram for measuring $M_{1,A}$ (left) and $M_{2,B}$ (right) for a two-qubit source emitting states $\rho$ with period $\tau$.  For both qubits from a pair the delay varies depending on the pair number in the sequence. In the case of measuring $M_{i,1}$ the qubits in arm $A$ ($B$) are delayed by  $2\tau$ or $\tau$ ($\tau$ or $0$), whereas for $M_{i,2}$ the qubits in arm $A$ ($B$) are delayed by $4\tau$, $\tau$, or $0$ ($\tau$ or $0$). The difficulty of implementing the delay in a real experiment depends on the value of $\tau$ and on technical aspects depending on the physical properties of the qubits. For a very small $\tau$ it can be difficult to provide fast enough switching of the delay, whereas for a very large $\tau$ coherent storage of qubits could be challenging.}
\end{figure}

\section{Detection of nonclassical correlations}

Up to this point we did not mention how to measure $Q$. The proposed measurement procedure is similar to the proposal for direct measurement of quantum discord introduced by Jin \etal  in \cite{Jin2012} and it resembles the so-called entanglement swapping \cite{Zukowski1993}.  In our case we need only the first two moments of the matrix $K_i$, and we need fewer copies of $\rho$. The first moment is given as $M_{1,A}=\mathrm{Tr}(\vec{x}\vec{x}^T+TT^T)$,  where we fixed without loss of generality $i=A$, and requires two copies of $\rho$ to be measured. This is because $M_{1,A}$ can be also expressed \cite{Jin2012} as
\begin{eqnarray}
M_{1,A}&=&\mathrm{Tr}[(U_{A_1A_2}\otimes V_{B_1B_2})(\rho_{A_1B_1}\otimes\rho_{A_2B_2})],\label{eq:m1}
\end{eqnarray}
where
\begin{subequations}
\begin{eqnarray}
U_{A_mA_n}&=&-4P^-_{A_mA_n}+I_{A_mA_n},\\ 
V_{B_mB_n} &=& U_{B_mB_n} + I_{B_mB_n}.\label{eq:U}
\end{eqnarray}
\end{subequations}
Both $U$ and $V$ are given in terms of singlet projections
$P^-_{i_mi_n}=(|\Psi^-\rangle\langle\Psi^-|)_{i_mi_n}$ and 
\begin{equation}
|\Psi^-\rangle_{i_mi_n}=\frac{1}{\sqrt{2}}(|\!\!\uparrow\rangle|\!\!\downarrow\rangle-|\!\!\downarrow\rangle|\!\!\uparrow\rangle)_{i_mi_n}
\end{equation}
with $i=A,B$ and $m,n=1,2$. However,  for measuring $M_{2,A}$ we need four copies $\rho$, since
\begin{eqnarray}
M_{2,A}&=&\mathrm{Tr}[(U_{A_1A_4}\otimes U_{A_2A_3}\otimes V_{B_1B_2}\otimes V_{B_3B_4})\cr\cr &&\times(\rho_{A_1B_1}\otimes\rho_{A_2B_2}\otimes\rho_{A_3B_3}\otimes\rho_{A_4B_4})].\label{eq:m2}
\end{eqnarray}

The operators $U_{A_mA_n}$ and $V_{B_mB_n}$ are  two-qubit operators acting on the respective subsystems. Having  four copies of $\rho$ each moment can be estimated in a single coincidence measurement. However, if six copies are available, the NW from Eq.~(\ref{eq:witnessA}) could be evaluated in a single measurement. We can however, use temporal separation between the copies (as shown in Figs.~\ref{fig:2} and~\ref{fig:3}) instead of spatial separation to investigate the quantum correlations of a two-qubit source of a constant frequency of $1/\tau$.

\section{Linear-optical implementation}

\begin{figure}
\includegraphics[scale=0.33]{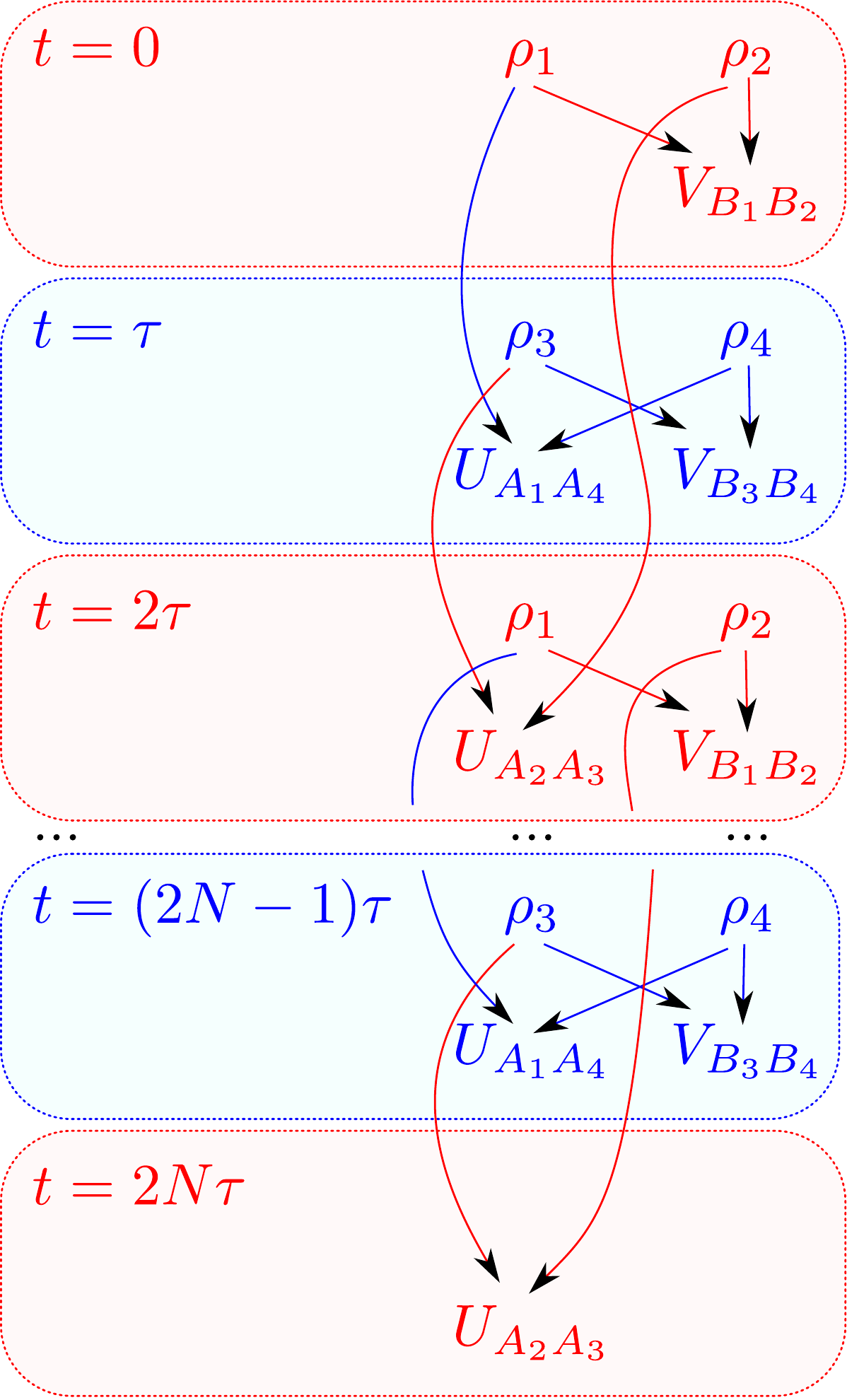}
\includegraphics[scale=0.33]{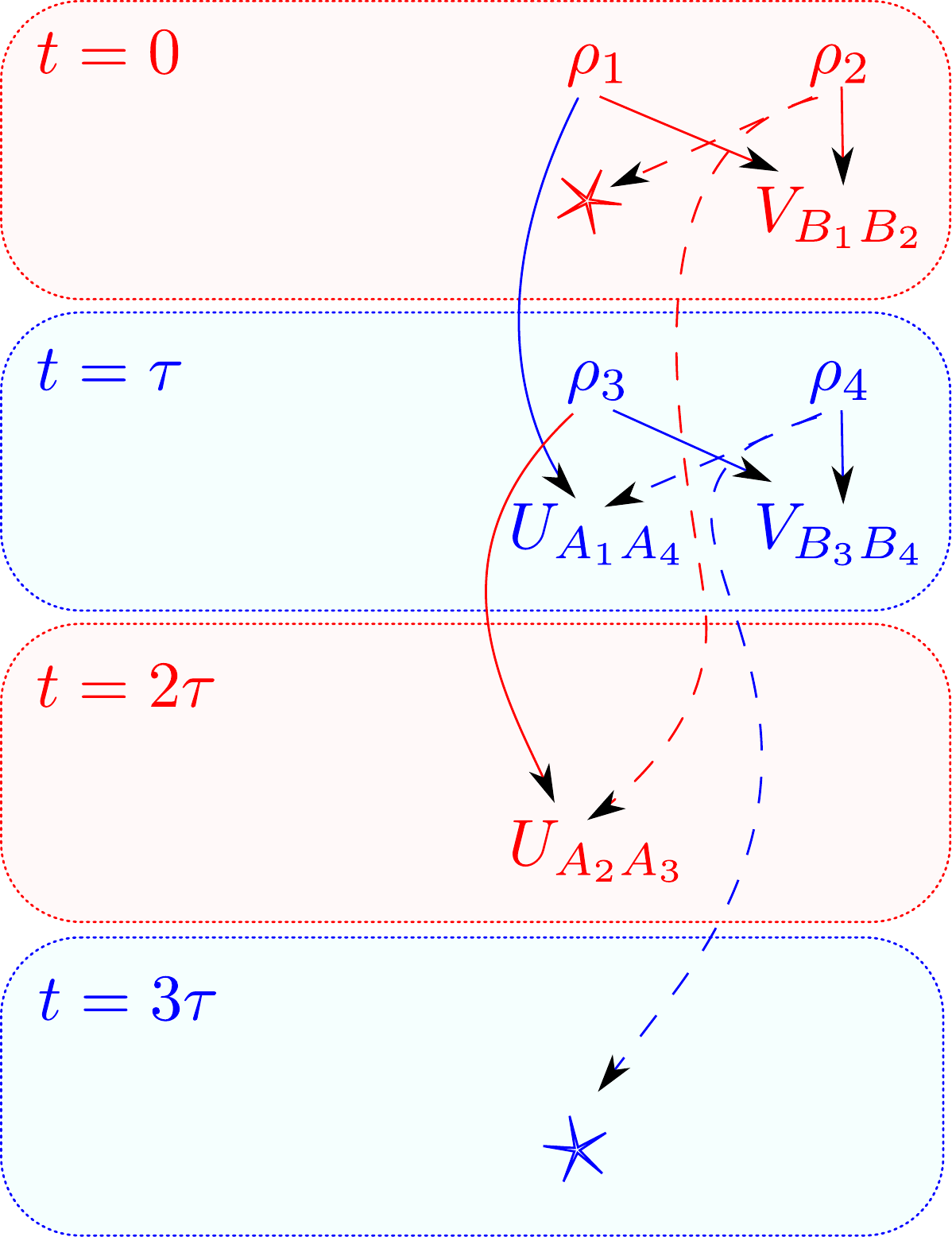}
\caption{\label{fig:3}(Color online) Diagram of measurement of $M_{2,A}$ using four photons (two two-qubit states $\rho$) emitted at intervals $\tau$. From left: $N$ iterations of the measurement procedure employing deterministic delay; a single iteration of the procedure in case of the probabilistic  delay. In the latter case the time required  for obtaining a single measurement outcome is twice as long as in the other case since in order to obtain the correct result one needs to ensure that photons left the delay lines. Thus,  the deterministic scheme is $2/p^2$ times more efficient than the probabilistic one, where $p$ is the success probability of the delay.  However, the deterministic scheme requires switching the duration of the delay with frequency $1/\tau$ equal to the repetition rate of the source. The measurement diagram for $M_{1,A}$ for a four-photon source is trivial since it does not require delays. }
\end{figure}

The  QDI from Eq.~(\ref{eq:witnessA}) can be measured in a linear-optical system where the qubit $z$-basis states $|\!\!\uparrow\,\equiv 1_H\rangle$ and  $|\!\!\downarrow\, \equiv 1_V \rangle$ are represented by single-photon horizontal  $|1_H\rangle$ and vertical $|1_V\rangle$ polarization states. Our proposal for the experimental setup consists of standard optical elements, i.e., 50:50 asymmetric beam splitters (BS) and photon detectors.  Since the qubits are polarization encoded it may seem surprising that we do not use polarization-dependent components as, e.g., polarizing BSs, however the polarization dependence that we take advantage of is induced in the bosonic commutation relations that are fulfilled by the BS transformed photons.  As shown in Fig.~\ref{fig:2} we need a two-photon source to conduct the experiment given that we are able to implement the correct sequence of delays for the corresponding photons. This task could can be very demanding and therefore we propose an alternative experiential  procedure using a four-photon source as shown in Fig.~\ref{fig:3}. High quality multiphoton sources of  are difficult to obtain (for a review see Ref.~\cite{Pan2012}) but implementing them is easier than providing the complex reliable and efficient system of delays
as described in Fig.~\ref{fig:2}.

\begin{figure}
\includegraphics[width=6cm]{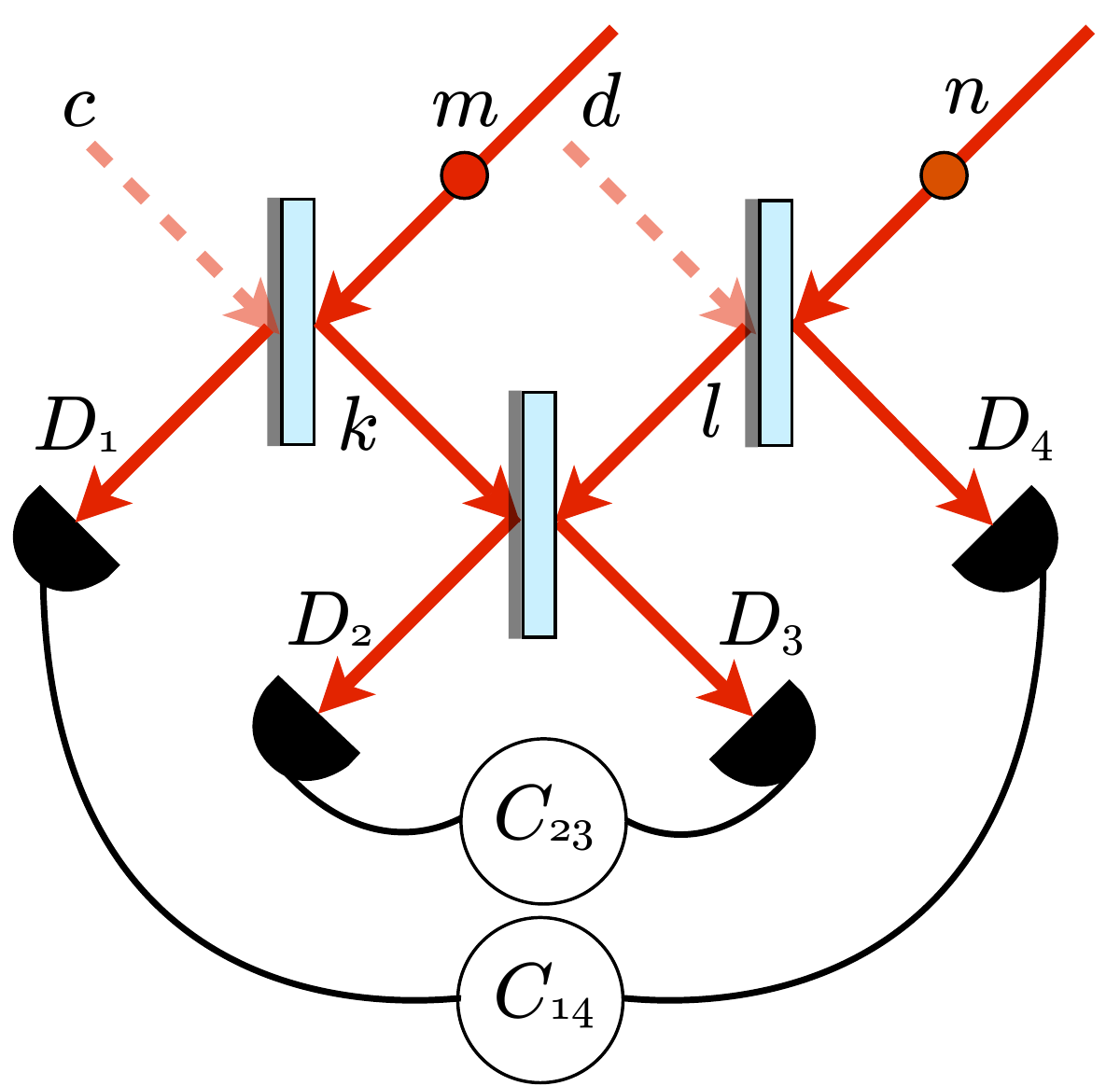}
\caption{\label{fig:4}(Color online)  Setup for linear-optical measurement of $U_{mn}$ or $V_{mn}$ observables, i.e., the $U/V$ box. The setup consists of 50:50 asymmetric beam splitters (BSs) and conventional detectors ($D$). To measure $U_{mn}$ ($V_{mn}$) a  value of $-4$ is assigned to  coincidence $C_{23}$ at $D_2$ and $D_3$ as it implies singlet projection [see Eq.~(\ref{eq:C23})] and value of 1 (2) if the coincidence $C_{14}$ implying $I_{mn}$ operation [see Eq.~(\ref{eq:C14})] is registered by $D_1$ and $D_4$.  The other combinations of detector clicks are irrelevant, thus we do not assign values to other detection events. The device works with probability $1/2$ because with this probability both photons reach unaltered the central BS or exit to detectors $D_1$ and $D_4$.}
\end{figure}

\subsection{Assembling a $U/V$ box}

The operators constituting moments $M_{1,A}$ and $M_{2,A}$ are $U_{A_mA_n}$ and $V_{B_mB_n}$. Each of the operators is a linear  combination of  the projections $P^-$ and identity operations $I$. Both of these fundamental operations can be implemented using linear optics (see Fig.~\ref{fig:4}) with polarization encoded qubits. Operation $I$ corresponds to detecting unaltered photons. The singlet projections $P^-$ is implemented by detecting a pair of photons in separate spatial modes, say 2 and 3, after they interacted on a BS assuming that both photons impinge on the BS from two separate modes, e.g.,  $m$ and $n$, as well. The asymmetric 50:50 BS transmits and reflects a photon with probability $1/2$, but the photon gets a $\pi$ phase shift only if it reflects from one side side of the BS (marked by a gray layer in Fig.~\ref{fig:4}). Note that the  BS transformation is unitary and is its own inverse, i.e., $\mathrm{BS} =\mathrm{BS}^\dagger$.

\subsubsection{Identity operator}

First let us show that the two-mode identity operator $I_{mn}$ can be implemented as a coincidence $C_{14}$ at detectors $D_1$ and $D_4$ shown in Fig.~\ref{fig:4}. This coincidence can be expressed as
\begin{equation}
C_{14}=\eta^2|1_1\rangle\langle1_1|\otimes|1_4\rangle\langle 1_4|,
\end{equation}
where $\eta$ is quantum efficiency of the detectors $\mathrm{D}_1$ and $\mathrm{D}_4$. We neglect the dark counts since depending on the wavelength and the detectors used they can be very rare, moreover the dark counts are uncorrelated and thus their simultaneous appearance is highly improbable. Once we trace back the photons detected by $C_{14}$ to the input modes $c,\,m,\,n,\,d$  of the setup from Fig.~\ref{fig:4} by performing the BS transformation  we obtain
\begin{eqnarray}
\nonumber
C'_{14} &=& r\left[|1_m\rangle\langle1_m|\otimes |1_n\rangle\langle1_n|  +|1_m\rangle\langle1_m|\otimes |1_d\rangle\langle1_d| \right.\\
&& \left. + |1_c\rangle\langle1_c|\otimes |1_n\rangle\langle1_n| + |1_c\rangle\langle1_c|\otimes|1_d\rangle\langle1_d|\right],
\end{eqnarray}
where $r = \eta^2/4$. Since no photons arrive form the vacuum modes $c$ and $d$ the $C_{14}$ operator effectively reads as
\begin{equation}
C'_{14} = r|1_m\rangle\langle1_m|_m\otimes|1_n\rangle\langle1_n|, 
\end{equation}
or in polarization basis where $|1_n\rangle\langle1_n| = ( |1_H\rangle\langle 1_H| + |1_V\rangle\langle 1_V|)_n = I_n$ it reads
\begin{equation}\label{eq:C14}
C'_{14} = r I_{m}\otimes I_{n} = r I_{mn}. 
\end{equation}
Thus, by performing coincidence detection after the two BSs we measure an operator proportional to $I_{mn}$, where the constant coefficient $r=\eta^2/4$ is known in advance.

\subsubsection{Singlet projection}

The analogous reasoning as in the case of the two-mode identity operator $I_mn$ can be applied to show that coincidence $C_{23}$ performs the $P^-_{mn} = (|\Psi^{\pm}\rangle\langle \Psi^-|)_{mn} $ projection. Let us start with the operator describing the coincidence at detectors  $D_2$ and $D_3$
\begin{equation}
C_{23}=\eta^2|1_2\rangle\langle1_2|\otimes|1_3\rangle\langle 1_3|
\end{equation}
which can be rewritten in the Bell basis 
\begin{subequations}
\begin{eqnarray}\label{eq:BellS}
|\Phi^{\pm}\rangle_{23}   &=&  \frac{1}{\sqrt{2}}\left(|1_H\rangle_2|1_H\rangle_3 \pm |1_V\rangle_2|1_V\rangle_3 \right) ,\\
|\Psi^{\pm}\rangle_{23}   &=&   \frac{1}{\sqrt{2}}\left(|1_H\rangle_2|1_V\rangle_3 \pm |1_V\rangle_2|1_H\rangle_3 \right)
\end{eqnarray} 
\end{subequations}
as
\begin{eqnarray}
\nonumber
C_{23}&=&\eta^2\left(|\Phi^+\rangle\langle\Phi^+|+|\Phi^-\rangle\langle\Phi^-| \right.\\
&&\left. + |\Psi^+\rangle\langle\Psi^+| + |\Psi^-\rangle\langle\Psi^-|\right)_{23}.
\end{eqnarray}
The Bell states are transformed by the BS as follows:
\begin{subequations}
\begin{eqnarray}
|\Phi^{\pm}\rangle_{23} &\to& \frac{1}{2}\left( |2_H\rangle_k |0\rangle_l + |0\rangle_k|2_H\rangle_l \right. \cr\cr && \left.\mp |2_V\rangle_k|0\rangle_l \pm |0\rangle_k|2_V\rangle_l \right),\\
|\Psi^{+}\rangle_{23} &\to& \frac{1}{\sqrt{2}}\left( |0\rangle_k|1_H,1_V\rangle_l - |1_H,1_V\rangle_k|0\rangle_l\right),\\
|\Psi^{-}\rangle_{23} &\to& -|\Psi^-\rangle_{kl}.
\end{eqnarray} 
\end{subequations}
Thus, the coincidence count after the action of the central BS from Fig.~\ref{fig:1} reads
\begin{eqnarray}
\nonumber
C'_{23}&=&\eta^2\left(|\Psi^-\rangle\langle\Psi^-|+\mathrm{BS}|\Psi^+\rangle\langle\Psi^+|\mathrm{BS}  \right.\\
&&\left. +\mathrm{BS}|\Phi^+\rangle\langle\Phi^+|\mathrm{BS}  
+\mathrm{BS}|\Phi^-\rangle\langle\Phi^-|\mathrm{BS} \right)_{kl}.
\end{eqnarray}
Note that all the BS transformed Bell states except the singlet state have two-photon components which are impossible to appear because there can be only one photon per mode $k$ and $l$. Thus, we can simplify the operator $C'_{23}$ describing the BS transformation followed by coincidence detection to 
\begin{equation}
C'_{23}=\eta^2(|\Psi^-\rangle\langle\Psi^-|)_{k,l}
\end{equation}
which in terms of the input modes reads
\begin{equation}\label{eq:C23}
C''_{23}=r(|\Psi^-\rangle\langle\Psi^-|)_{mn} = r P^-_{mn},
\end{equation}
where the terms describing the detection of photons emerging from vacuum modes $c$ and $d$ were not taken into account.

\subsubsection{Measuring $U$ and $V$}

The operators $U_{mn}$ and $V_{mn}$ are linear combinations of $C'_{14}$ and $C''_{23}$ of the following form
\begin{subequations}
\begin{eqnarray}
U_{mn} &=& \frac{1}{r} (C'_{14} - 4 C''_{23}),\\
V_{mn} &=& \frac{1}{r} (2C'_{14} - 4 C''_{23}).
\end{eqnarray}
\end{subequations}
The factor of $1/r=4/\eta^2 > 1$ compensates for the fact that $C'_{14}$ or $C''_{23}$ are both measured probabilistically with a success rate of $r = \eta^2/4$, but the total success rate $R$ of the $U/V$ setup is $R =2r$.  The probabilistic nature of the setup reduces uniformly the number of all the coincidence counts by a factor of $r=\eta^2/4$. Thus, we can express the average values of $\langle U_{mn} \rangle$ and $\langle V_{mn} \rangle$ as
\begin{subequations}
\begin{eqnarray}
\langle U_{mn} \rangle &=& 1-4\frac{N_{23}}{N_{14}},\\
\langle V_{mn} \rangle &=& 2-4\frac{N_{23}}{N_{14}},
\end{eqnarray}
\end{subequations}
where $N_{14}$ is the number of registered coincidences $C_{14}$ and $N_{23}$  is the number of registered coincidences $C_{23}$. 

\subsection{Measuring $M_1$ and $M_2$}

\begin{figure}
\includegraphics[width=8cm]{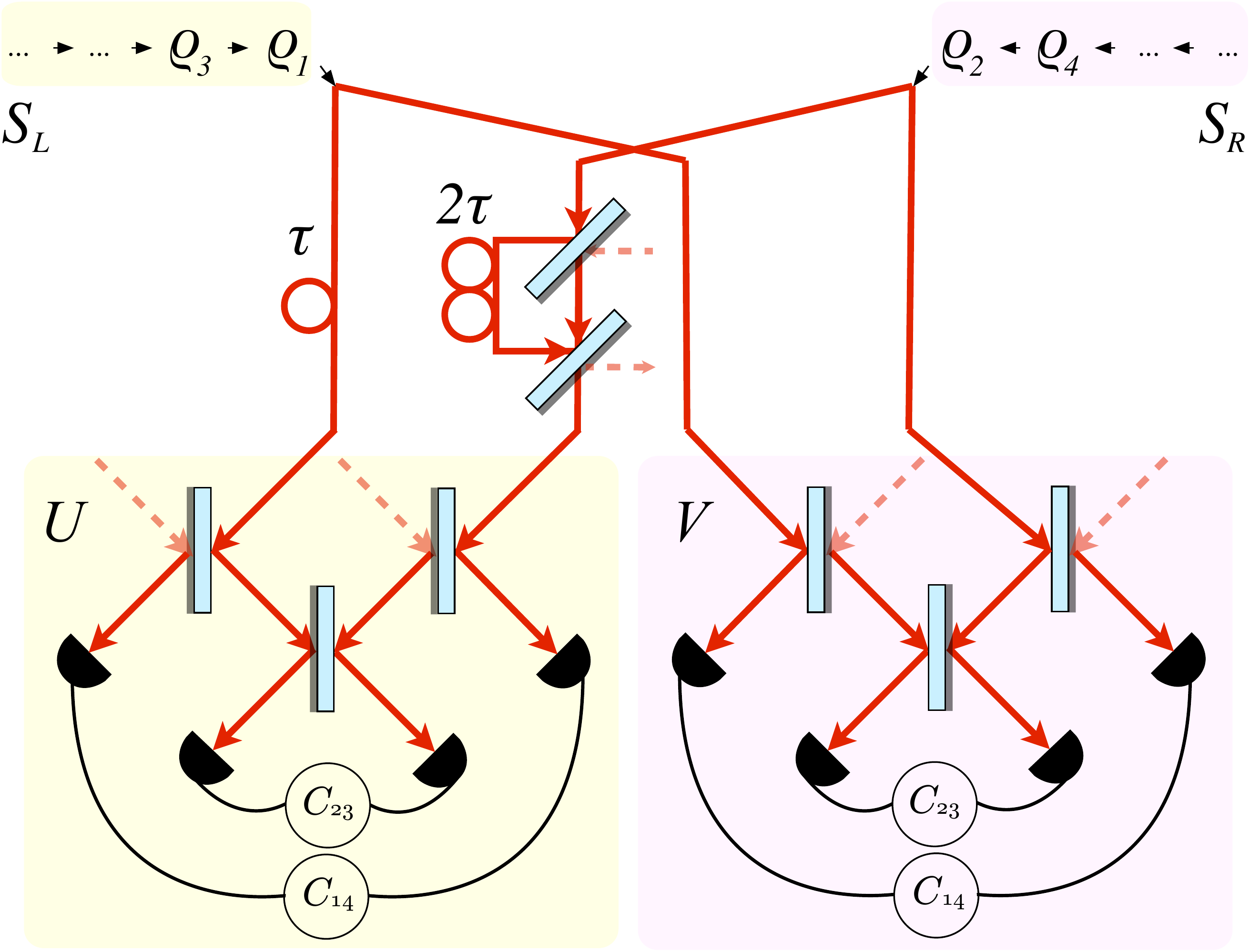}
\caption{\label{fig:5}(Color online) Optical setup for measuring $M_2$
using the probabilistic scheme from Fig.~\ref{fig:3} with $p=1/4$. When the delay lines are removed, the setup measures $M_1$.  At times  $t=0$ and $t=\tau$ two photon pairs $\rho$ enter the setup. When $t=2\tau,\,3\tau$ no new photons appear from sources $S_1$ and $S_2$ and only coincidences in the $U$ part of the setup are witnessed. The photons are delayed before entering the $U$ part. One of the photons is always delayed by $\tau$ and the second one randomly by $2\tau$ or $0$, however the photon  with probability $0.5$ can leak out of the system. In the successful cases  $\rho_{A_1B_1}$ and $\rho_{A_2B_2}$ enter the setup at $t=0$ and $V_{B_1B_2}$ is measured. At the same time photons in the left part are delayed, the first from the left by $\tau$ and the second by $2\tau$ (with probability  $1/4$), so at $t=0$ there is no photon detected in $U$. Photon pairs $\rho_{A_3B_3}$ and $\rho_{A_4B_4}$ enter the setup at $t=\tau$ and $V_{B_3B_4}$ is measured without delay. In the left part, the first photon is delayed by $\tau$, but the second one is not affected (with probability $1/4$). At the same time $U_{A_1A_4}$ is measured. Since there are no photons added at $t=2\tau$ only $U_{A_2A_3}$ is measured (for a photon provided at $t=0$ and delayed by $2\tau$ and a photon provided at $t=\tau$ delayed by $\tau$). The success rate of the delay procedure is $p^2=1/16$. The whole sequence is repeated until a good estimate of $M_2$ is obtained, e.g., we reach $N_2=10^3$ successful iterations. No photons enter the setup at $t=3\tau$ to not affect the next measurement iteration.}
\end{figure}

Estimating the averages of products of the $U$ and $V$ operators is a bit more involved since it requires using more than one $U/V$ block shown in Fig.~\ref{fig:4}. Our experimental proposal based on the probabilistic scheme from Fig.~\ref{fig:3} shown in Fig.~\ref{fig:5} uses two such blocks. Let us assume that we perform $N$ iterations of the experiment for $n=1,2...,N$. For each iteration the $U/V$ box provides us with one of three values $u_{mn} \in \{0,1,-4\}$ for measuring $U_{mn}$ and $v_{mn} \in \{0,2,-4\}$ for measuring $V_{mn}$. The there values ($\{0,1,-4\}$ or $\{0,2,-4\}$) correspond to not observing coincidence $C_{14}$ or $C_{23}$, observing coincidence $C_{14}$, and observing coincidence $C_{23}$, correspondingly. Thus, for measuring $\langle M_{1,A} \rangle = \langle U_{A_1A_2}\otimes V_{B_1B_2}\rangle$ the $n$'th measurement outcome is a product of two numbers and reads $a_n = (u_{A_1A_2}v_{B_1B_2})_n$, whereas for $\langle M_{2,A} \rangle = \langle U_{A_1A_4}\otimes U_{A_2A_3}\otimes V_{B_1B_2}\otimes V_{B_3B_4}\rangle$ it is a product of four numbers, i.e., $b_n = (u_{A_1A_4}u_{A_2A_3} v_{B_1B_2}v_{B_3B_4})_n$. We can express the expectation values as
\begin{subequations}
\begin{eqnarray}
\langle M_{1,A} \rangle &=&\frac{1}{N_1}\sum_{n=1}^N a_n,\\
\langle M_{2,A} \rangle &=&\frac{1}{N_2}\sum_{n=1}^N b_n,
\end{eqnarray}
\end{subequations}
where the number of events where $I^{\otimes 2}$ or $I^{\otimes 4}$ was measured  is $N_1 = \sum_{n=1}^N \delta_{2,a_n}= r^2N$ and $N_2 =  \sum_{n=1}^N \delta_{4,b_n}= p^2r^4N$, correspondingly, where $\delta$ is the Kronecker's $\delta$ and $p$ is the success probability of the delay. 

\subsection{Experimental challenges}

\begin{figure}
\includegraphics[width=7cm]{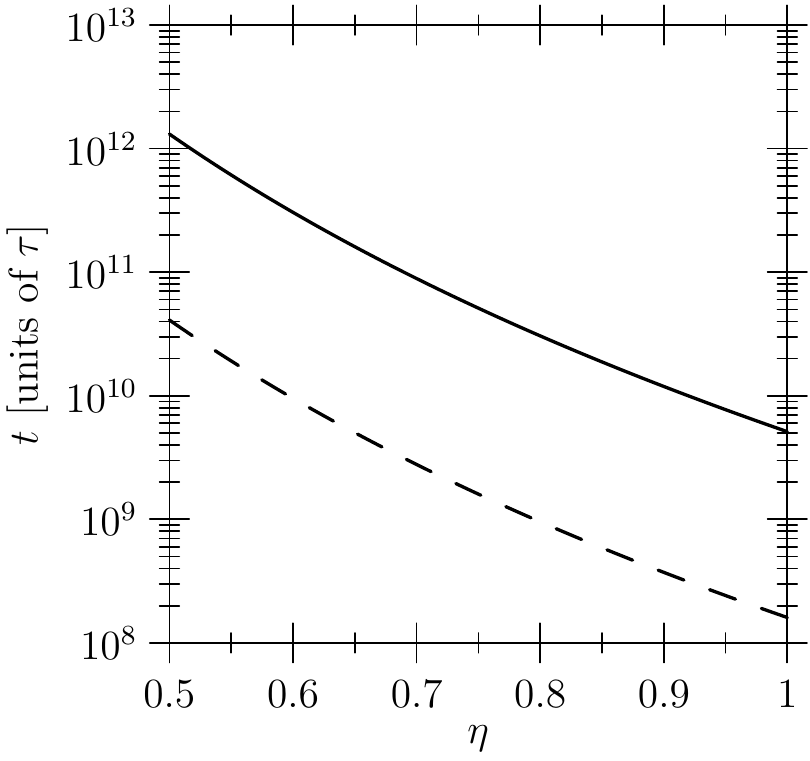}
\caption{\label{fig:6} Time $t$ in units of $\tau$ (the inverse of time between generation of the two two-photon states) needed for performing $10^3$ successful measurements of $M_1$ and $M_2$ versus the detection efficiency $\eta$. The success rate of generating two pairs of photons from a single pump pulse was assumed to be $1/100$.  The solid curve corresponds to the probabilistic delay scheme whereas the dashed one corresponds to the deterministic delay scheme. The repetition rate of the pump can be set to $1/\tau=20\,\mathrm{MHz}$ ($\tau = 50\,\mathrm{ns}$). Te repetition rate is limited by the dead time of the detectors (typically about $50\,\mathrm{ns}$). Measuring both moments with $10^3$ iterations in a time of order of $1\,\mathrm{s}$ would be possible for perfect detectors and deterministic delay scheme. In a realistic case of $\eta = 0.75$ and probabilistic delay the whole measurement would take less than $3\,\mathrm{min}$. Note, however, that $N=100$ estimates for $M_1$ and $M_2$ would be available in a 10 times shorter time.}
\end{figure}

Note that for the case of measuring $M_{1,A}$ and  $M_{2,A}$ we lose $(1-R^2)N$ and $(1-p^2R^4)N$ photons, correspondingly. In spite of losing many photons measuring $Q_i$ is expected to be faster than full 2-qubit \textit{quantum tomography} (QT) which requires rotating a number of polarization plates providing 16 measurement configurations, where each rotation takes a few seconds. During this time for small $\tau$ the measurement setup from Fig.~\ref{fig:5} would accumulate enough data to estimate $Q_i$.  This assessment, however, strongly depends on the efficiency of the detectors $\eta$ (see Fig.~\ref{fig:6}). 

\subsubsection{Speed of the measurement}

The main difficulty in implementing the setup outlined in Fig.~\ref{fig:5} is working with two two-photon sources $S_R$ and $S_L$. The photon pairs should exhibit quantum correlations, thus should be produced, e.g., in spontaneous parametric  down-conversion of light of a certain degree of depolarization. Since the pairs are produced by a random process, the coincidence rate in our experimental proposal can be very low. However, as we demonstrate below, it is not always the case. There are of course further challenges. Depending on the approach of producing the input states, one may face the problem of indistinguishability of the photon pairs, which causes the results to differ from the expected one, but methods of circumventing this effect can be found, e.g., in \cite{Pan2012}. Moreover, the theoretical prediction is reached if all the pairs perfectly overlap on the corresponding BSs. This condition is easy to be satisfied for one pair of photons, however the amount of work increases with the number of photons. To achieve this all the optical elements would require active stabilization. The success rate of the measurement of $M_2$ from Fig.~\ref{fig:5} is $p_2=p^2 R^4$ since it requires applying probabilistic delay for measuring $U$. When all the delay lines in Fig.~\ref{fig:5} are removed, the setup and can be used for direct measurement of $M_1$ with success rate of $p_1=R^2$. The success rates should also incorporate the probabilities of generating photon pairs at $t=n\tau$. In two-crystal type-I geometry (so-called Kwiat source \cite{Kwiat1999}), the probability of generating an entangled photon pair from a single pump pulse is about $1/10$. Thus, creating two pairs independently occurs with probability $1/100$ and the effective success probabilities read $p_1 = R^2/100$ and $p_2 = p_1^2$ for $M_1$ and $M_2$, correspondingly. By taking into account  the pauses in providing two pairs of photons at $t=2\tau, 3\tau$ (we only allow every second two pulses from the pulse train to enter the setup by using a pulse picker) the efficiency of the setup for measuring $M_2$ drops by a factor of $1/2$. Thus, the useful coincidence count rate for $M_2$ would be limited to $p_2=p_1^2/2$ of the brightness of the source. The repetition rate of the source ($1/\tau$) is limited by the dead time of the detectors (about 50\,ns) to about 20\,MHz. Thus, for realistic detector efficiency of $\eta=0.75$ (Perkin-Elmer single photon counting modules operating at wavelength of 700\,nm), the number of successful $M_2$ measurements per second would reach  $p_2\times 20\,\mathrm{MHz}=6.25\,\mathrm{Hz}$ and $p_1\times 20\,\mathrm{MHz}=15.8\,\mathrm{kHz}$ for $M_1$. These numbers would be further reduced by a few percent by unavoidable imperfections of the setup. Nevertheless, the final expected number of detection events for $M_2$  is of order of $6$ coincidences per second which would allow us to measure $M_2$ in less than three minutes. Measuring $M_1$ would be much faster because of the simpler structure of coincidences, no pulse picking, and lack of probabilistic delay. The deterministic setup for measuring $M_2$ outlined  in Fig.~\ref{fig:3} would also not suffer from such a low coincidence rate as the setup shown in Fig.~\ref{fig:5}. However, the later alternative is more feasible and can be successfully implemented, e.g., in our laboratory. 

\subsubsection{Robustness of the setup}

\begin{figure}
\includegraphics[width=7cm]{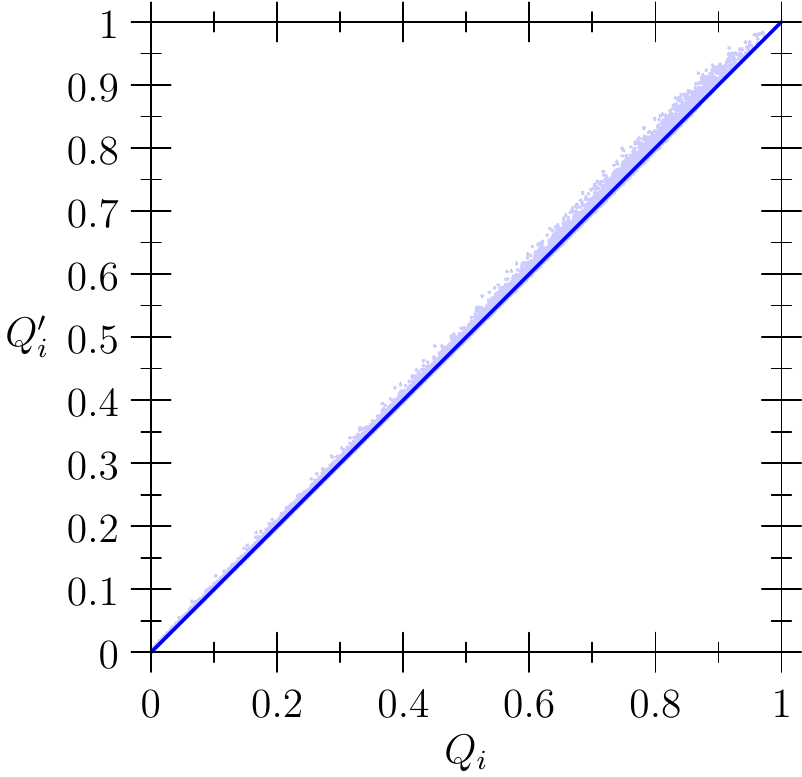}
\caption{\label{fig:7}  The measured value of $Q'_A$ for the two sources $S_L$ and $S_R$ producing pairs of photons in states $\rho^L$ and $\rho^R$  versus the precise $Q_A$ value for $\rho = (\rho^L+\rho^R)/2$. The blue data points are Monte Carlo results  for $10^6$ randomly generated $\rho^R$ and $\rho^L$ states satisfying $F(\rho^L,\rho^R)\geq 0.90$. The blue line corresponds indicates $Q'_A = Q_A$. }
\end{figure}

For the purpose of our experiment we need two two-photon sources $S_L$ and $S_R$ in Fig.~\ref{fig:5} producing photon pairs in the same state $\rho$. However, the sources can produce slightly different states $\rho^L$ and $\rho^R$ for  $S_L$ and $S_R$, correspondingly. In this case we assume that our state $\rho$ is given as $\rho = (\rho^L + \rho^R)/2$. If the states $\rho^L$ and $\rho^R$ are not identical  the fidelity  \cite{Jozsa94} of these two states
\begin{equation}
F(\rho^L,\rho^R) =\left\lbrace \mathrm{Tr}\left[ \left({\sqrt{\rho^L}\rho^R\sqrt{\rho^L}}\right)^{1/2}\right]\right\rbrace^2
\end{equation}
is less than $1$ and the measured moments read
\begin{eqnarray}
M'_{1,A}&=&\mathrm{Tr}[(U_{A_1A_2}\otimes V_{B_1B_2})\times(\rho^L_{A_1B_1}\otimes\rho^R_{A_2B_2})],\label{eq:Dm1}
\end{eqnarray}
and
\begin{eqnarray}
M'_{2,A}&=&\mathrm{Tr}[(U_{A_1A_4}\otimes U_{A_2A_3}\otimes V_{B_1B_2}\otimes V_{B_3B_4})\cr\cr && \times(\rho^L_{A_1B_1}\otimes\rho^R_{A_2B_2}\otimes\rho^L_{A_3B_3}\otimes\rho^R_{A_4B_4})].\label{eq:Dm2}
\end{eqnarray}
Using the moments (\ref{eq:Dm1}) and (\ref{eq:Dm2}) for calculating $Q_A$ we obtain $Q'_A=Q(M'_{1,A},M'_{2,A})$ which varies with fidelity  $F(\rho^L,\rho^R)$ of the sources $S_L$ and $S_R$. The value of $Q'_A$ can be successfully used to approximate $Q_A(\rho)$, where $\rho = (\rho^L + \rho^R)/2$. For reasonably large fidelity values $F(\rho^L,\rho^R)\geq 0.90$ this approximation is expected to be very accurate since for the $10^6$ Monte Carlo generated states $\rho^L$ and $\rho^R$ shown in Fig.~\ref{fig:7} the largest module of the difference $Q'_A-Q_A$ is smaller than 0.05, i.e.,  $|Q'_A-Q_A|<0.05$. Obtaining fidelity above $F=0.90$ should not be difficult, thus, the setup can be considered robust to imperfections of the sources.

\section{Conlusions}

We demonstrated that $Q$ of Girolami and Adesso can be efficiently measured in linear-optical system utilizing two sources of two-photon states.  Due to its properties the QDI which does not require any prior knowledge about the investigated state $Q$ is more powerful than the other  recently reported NWs \cite{Rahimi2010, Auccaise2011, Passante2011, Maziero2012, Aguilar2012}. The QDI and the NW introduced by Zhang \etal  in \cite{Yu2011} both require four copies of the input state, but the latter one is not appropriate for optical implementation. Our approach is less complex than full QT, since it only requires two measurements, if four copies of $\rho$ are available [either spatially (see Fig.~\ref{fig:2}) or time separated (see Fig.~\ref{fig:3})], instead of 16 for the QT \cite{Xu2010,Xu2011,Soares2010}. Furthermore, the proposed measurement setup is expected to be faster than the direct QT, since it would provide the outcome within a minute. Let us note that if the one had access to six copies of $\rho$, the QID could be estimated in a single measurement in contrast to the simpler experiment described in \cite{Aguilar2012}, where the NW was not universal and its estimation involved measuring three quantities. Our approach can be extended (by using larger number of copies) to directly measure GQD as discussed in Ref.~\cite{Jin2012}. Moreover, our results can help to establish discord related NWs based on matrices of moments for optical fields (studied previously in the context of  entanglement~\cite{Shchukin05} and other nonclassical phenomena~\cite{Vogel08}). Finally, let us note that the discussed setup can measure moments $M_1$ and $M_2$ which can constitute other QDIs or NWs (see Ref.~\cite{Girolami12}), thus it can be used for investigating various quantum properties of light in a way that was previously available only for the NMR platform.

\begin{acknowledgments}

The authors thank A. Miranowicz for fruitful discussions. The authors gratefully acknowledge the support by the Operational Program Research and Development for Innovations -- European Regional Development Fund (Project No.~CZ.1.05/2.1.00/03.0058)  and the  Operational Program Education for Competitiveness - European Social Fund (Projects No. CZ.1.07/2.3.00/20.0017, No. CZ.1.07/2.3,00/20,0058, No. CZ.1.07/2.3.00/30.0004, and No.  CZ.1.07/2.3.00/30.0041) of the Ministry of Education, Youth and Sports of the Czech Republic. The authors also acknowledge  support by the Internal Grant Agency of Palacky University in Olomouc (project No. PrF\_2012\_003). K.B. was supported by Grant No. DEC-2011/03/B/ST2/01903 of the Polish National Science Centre.
\end{acknowledgments}



\end{document}